\documentstyle[12pt,epsf,epsfig]{article}  
\def\epsi{\epsilon}

\def\L{{\cal L}}
\def\O{{\cal O}}

\def\li2{{\rm Li}_2}

\def\roughly#1{\,\,\raise.3ex\hbox{$#1$\kern-.75em\lower1ex\hbox{$\sim$}}\,\,}

\def \lsim{\mathrel{\vcenter
     {\hbox{$<$}\nointerlineskip\hbox{$\sim$}}}}
\def \gsim{\mathrel{\vcenter
     {\hbox{$>$}\nointerlineskip\hbox{$\sim$}}}}

\def\fo{\hbox{{1}\kern-.25em\hbox{l}}}

\def\bea{\begin{eqnarray}}
\def\eea{\end{eqnarray}}
\def\beq{\begin{equation}}
\def\eeq{\end{equation}}
\def\eq{\end{equation}}
\def\to{\rightarrow}

\def\bsg{\ifmmode B\to X_s\gamma\else $B\to X_s\gamma$\fi}
\def\bsll{\ifmmode B\to X_s\ell^+\ell^-\else $B\to X_s\ell^+\ell^-$\fi}
\def\bstt{\ifmmode B\to X_s\tau^+\tau^-\else $B\to X_s\tau^+\tau^-$\fi}
\def\shat{\ifmmode \hat{s}\else $\hat{s}$\fi}

\newcommand{\newc}{\newcommand}

\newc{\lcal}{\int {\cal L}dt}
 
\newc{\LSP}{{\chi^0_1}}
\newc{\stauR}{{\tilde \tau_R}}
\newc{\stau}{{\tilde \tau_1}}
\newc{\mstop}{m_{\tilde{t}}}
\newc{\mHpm}{m_{H^\pm}}
\newc{\ie}{{\it i.e.}}          
\newc{\etal}{{\it et al.}}
\newc{\eg}{{\it e.g.}}          
\newc{\kev}{\hbox{\rm\,keV}}            
\newc{\mev}{\hbox{\rm\,MeV}}            
\newc{\gev}{\hbox{\rm\,GeV}}            
\newc{\tev}{\hbox{\rm\,TeV}}
\newc{\xpb}{\hbox{\rm\, pb}}
\newc{\xfb}{\hbox{\rm\, fb}}

\newc{\mtop}{m_t}
\newc{\mbot}{m_b}
\newc{\mz}{m_Z}
\newc{\mw}{M_W}
\newc{\alphasmz}{\alpha_s(m_Z^2)}
\newc{\swsq}{\sin^2\theta_W}
\newc{\tw}{\tan\theta_W}
\newc{\cw}{\cos\theta_W}
\newc{\sw}{\sin\theta_W}
\newc{\BR}{\hbox{\rm BR}}
\newc{\zbb}{Z\to b\bar}
\newc{\Gb}{\Gamma (Z\to b\bar b)}
\newc{\Gh}{\Gamma (Z\to \hbox{\rm hadrons})}
\newc{\rbsm}{R_b^\hbox{\rm sm}}
\newc{\rbsusy}{R_b^\hbox{\rm susy}}
\newc{\drb}{\delta R_b}

\newc{\sgn}{\mbox{sgn}}
\newc{\tbeta}{\tan\beta}
\newc{\uL}{{\tilde u_L}}
\newc{\uR}{{\tilde u_R}}
\newc{\cL}{{\tilde c_L}}
\newc{\cR}{{\tilde c_R}}
\newc{\tL}{{\tilde t_L}}
\newc{\tR}{{\tilde t_R}}
\newc{\dL}{{\tilde d_L}}
\newc{\dR}{{\tilde d_R}}
\newc{\sL}{{\tilde s_L}}
\newc{\sR}{{\tilde s_R}}
\newc{\bL}{{\tilde b_L}}
\newc{\bR}{{\tilde b_R}}
\newc{\eL}{{\tilde e_L}}
\newc{\eR}{{\tilde e_R}}
\newc{\mhp}{m_{H^\pm}}
\newc{\mhalf}{m_{1/2}}
\newc{\emt}{{e/\mu /\tau}}

\def\lappeq{\mathrel{\rlap{\raise.5ex\hbox{$<$}}
{\lower.5ex\hbox{$\sim$}}}}
\def\gappeq{\mathrel{\rlap {\raise.5ex\hbox{$>$}}
{\lower.5ex\hbox{$\sim$}}}}

\newcommand{\drawsquare}[2]{\hbox{%
\rule{#2pt}{#1pt}\hskip-#2pt
\rule{#1pt}{#2pt}\hskip-#1pt
\rule[#1pt]{#1pt}{#2pt}}\rule[#1pt]{#2pt}{#2pt}\hskip-#2pt
\rule{#2pt}{#1pt}}

\newcommand{\Dal}{\drawsquare{7}{0.6}}

\begin{document}

\baselineskip=18pt



 


\vspace{20pt}
\font\cmss=cmss10 \font\cmsss=cmss10 at 7pt 

\hfill SNS-PH/00-21 

\hfill Bicocca-FT/00/25 \vskip .1in 
\hfill hep-th/0012248

\hfill

\vspace{20pt}

\begin{center}
{\Large \textbf
{Comments on the Holographic\\
Picture of the Randall-Sundrum Model}}
\end{center}

\vspace{6pt}

\begin{center}
\textsl{R. Rattazzi $^a$ and A.
Zaffaroni $^{b}$} \vspace{20pt}

$^{a}$\textit{I.N.F.N. and Scuola Normale Superiore, Pisa}

\textit{$^{b}$ I.N.F.N. and Universit\`{a} di Milano-Bicocca, Milano}
\end{center}

\vspace{12pt}

\begin{center}
\textbf{Abstract }
\end{center}

\vspace{4pt} {\small \noindent 
We discuss some issues about the holographic interpretation
of the compact Randall-Sundrum model, which is conjectured to be dual
to a 4d field theory with non-linearly realized conformal symmetry. 
We make several checks of this conjecture. In particular, we show
that the radion couples conformally to a background 4d metric.
We also discuss the interpretation of
the Goldberger-Wise mechanism for stabilizing the radion.
We consider situations where the electroweak breaking stabilizes
the radion and we discuss the issue of natural conservation of flavor
quantum numbers.}

\vfill\eject 
\noindent


\section{\large Introduction}

The observed hierarchy between the Planck and Fermi scales
suggests that there may be a more fundamental theory replacing
the Standard Model (SM) just above the latter scale. Within this 
theory the hierarchy of scales should arise as a natural
dynamical fact. This is what happens 
in technicolor and  supersymmetry.
Recently it was realized \cite{add,aadd,all} that extra space dimensions 
can offer a  new view 
point on the hierarchy problem, where the Fermi scale is essentially 
the fundamental quantum gravity scale  and the weakness of 4d-gravity is
 explained by a very large volume of compactification. 
Later Randall and Sundrum (RS) \cite{rs}
have pointed out that a small and warped
extra dimension can elegantly solve the hierarchy problem. The basic 
idea is that
we live on a 3-brane which is deep inside a five dimensional 
gravitational field\cite{rs}. Because of this, all the 
dimensionful parameters that describe
the SM are redshifted with respect to possibly similar
 branes located at other points. In the RS set up 
(called from now on RSI) the fifth dimension
is an $S_1/Z_2$ orbifold with locally $AdS$5 geometry and
bordered by two 3-branes with equal and opposite tensions. The $AdS$ warp 
factor is exponential in the fifth coordinate so that the 
energy  scales on the negative tension brane are also
exponentially redshifted. If we live on this brane
then we can naturally understand the small ratio $M_W/M_P$ as due
to an exponential red shift. Another way to view this is that the zero
mode corresponding to the 4d graviton is localized at
the positive brane (Planck brane) and has but a small overlap with the
negative brane (TeV brane). 

In the RSI model the graviton KK modes become  strongly coupled among
themselves and to the TeV brane just a little bit over a TeV, as expected in 
a theory where the fundamental quantum gravity scale is $\sim$TeV.
On the other hand, an observer doing local measurements 
on the Planck brane would not see strong coupling phenomena at energies
below $\sim M_P=10^{19}$ GeV. One way to see this is that the  light
KK graviton modes couple very weakly to the Planck brane. The fact that 
observers at different points enter the strong regime at different energies
can be confusing. What is the genuine scale of quantum gravity in this 
model? One way to address this question is to forget for a moment the
hierarchy problem and send the negative brane all the way to the 
$AdS$ horizon. Now spacetime is $AdS$5 wit the boundary region
truncated and replaced by the Planck brane. Even though the fifth
dimension is non-compact there still is a 4d graviton bound to the 
Planck brane. This is the RSII model\cite{rs2}. Based on the $AdS$/CFT 
correspondence \cite{maldacena}
is has been pointed out in \cite{wittensb,gubser} 
that the RSII can be considered
 just  a dual description
of a strongly coupled 4d CFT where 4d gravity has been weakly gauged. 
In the 5d picture the 
weak gauging of 4d gravity is due to the presence of a truncation (Planck
brane) removing the boundary of $AdS$5. In the limit were the Planck brane 
is moved to the boundary, 4d gravity decouples and one gets back just
a four dimensional CFT. 

This holographic duality is also useful to improve our intuition on
the RSI model and on the way it solves the hierarchy problem 
\cite{verlinde1,verlinde2, apr}. 
>From the holographic view point one can interpret the fifth coordinate 
as the renormalization scale in the 4d theory. The fact that $AdS$5 does 
not continue all the way down to the horizon but rather abruptly ends 
at the TeV brane corresponds to breakdown of conformal invariance below a TeV.
The 4d theory localized on this brane is 
naturally  interpreted as the low energy
end point of RG evolution, {\it i.e} the low energy description of
a strongly coupled (quasi)-CFT. 
Similarly, the Planck brane determines
the UV boundary conditions while the fields that live on it 
represent some hidden sector coupled to the CFT only via $1/M_P$
suppressed operators. 
One may expect that in a more fundamental
(string theory) description of the RSI set up, the metric singularity
 at the Planck and TeV branes will be smoothly resolved \cite{verlinde1,
verlinde2}.
So in the end the holographic interpretation of RSI is a 4d strongly
coupled theory which stays essentially conformally invariant all the way
down to the weak scale, below which it is effectively described by the 
Standard Model. This view point has been emphasized in \cite{apr}
with various arguments. 
In this letter we will elaborate a little more on the 
interpretation of the RSI, by showing that the graviton-radion system
is conformally invariant.
We will also discuss the stability
of $M_W/M_P$ from the 4d point of view, which is
crucial to decide the relevance of this scenario to the
hierarchy problem. We will discuss the 4d interpretation of the 
Golberger-Wise (GW)
mechanism for stabilizing the extra-dimension while obtaining a natural
hierarchy\cite{gw}. We will argue that from
the 4d point of view the GW mechanism is pretty generic. 
Similar arguments are presented in \cite{apr}.
Furthermore we will consider more general situations where the
electroweak breaking itself stabilizes the radion. 
Thinking of RSI as just a 4d field theory, rather
than being reductive, we believe helps putting more theoretical
requirements and therefore making more predictions.
We will illustrate this by considering the issue of natural 
conservation of flavor (and baryon and lepton) quantum numbers.

In Section 3 and 4, there is  overlap with \cite{apr}.
Some of the material and the motivations for this paper indeed 
resulted from discussions with the authors of \cite{apr}.
\section{\large GW radius stabilization: 5d picture}

We briefly establish our notation. 
Consider a bulk five-dimensional theory
\beq
{\cal L}=\int d^5x \sqrt{-g}(2M^3R-\Lambda_5)
\label{lag}\eeq
The metric for RSI is given by
\beq
ds^2=\frac{L^2}{z^2} \left( dx_\mu dx^\mu +dz^2\right)
\label{e1}
\eeq
where $L=1/k$ is the $AdS$ radius and the orbifold extends from $z=z_0=L$ 
(Planck brane) to $z=z_1$ (TeV brane). Using as fifth coordinate
$y=L \ln (z/L)$ we recover the usual parameterization with an exponential
warp factor. $z$ is the natural coordinate to discuss holography:
$1/z$ can be interpreted as the renormalization scale. So the presence
of the Planck brane at $z=z_0$ specifies that the 4d theory has a UV
cut-off $\mu_0= 1/z_0$, while $\mu_1= 1/z_1$ 
represents the IR cut-off, the Fermi scale.
The four dimensional Planck scale is given by
\beq
M_P^2=M^3 L^3\left (\frac{1}{z_0^2}-\frac{1}{z_1^2}\right)
=(ML)^3(\mu_0^2-\mu_1^2 ).
\eeq
The last expression is suited for a 4d interpretation \cite{gubser,apr}:
$c=4\pi^2(ML)^3$ represents the central charge of the CFT 
\cite{witten,hs} and $M_P$
is determined by a quadratically divergent quantum correction.
In the limit $\mu_0\to \infty$ 4d gravity decouples \cite{apr} and one
recovers a pure CFT as in the $AdS$/CFT correspondence.

In the original RS paper the fifth dimension is an exactly flat direction.
This is achieved by a suitable tuning of the model parameters. The 
tensions of the Planck and TeV branes are respectively given
by $T_0=+24M^3 k$ and $T_1=-24M^3k$, while the bulk cosmological constant
is $\Lambda_5=-24M^3k^2$. One of the tunings ensures that the effective
4d cosmological  constant vanishes, while the other is equivalent to require
 that the radius be a flat direction \cite{gw}\footnote{Basically this can also be seen as flatness of
the two moduli corresponding to the brane positions $z_0$ and $z_1$. Notice
however that only one combination, say $z_0/z_1$ corresponds to a 4d scalar.
The other, say $z_0$, represents the conformal factor of the 4d metric.}.
 The latter requirement
is however not needed and actually phenomenologically unacceptable.
Moreover, in order to argue that the model solves the hierarchy problem,
a dynamical mechanism that naturally selects $z_1\gg z_0$ must be found.
Goldberger and Wise have shown that simply a bulk scalar field $\phi$
can do the job. The action of such field is
\beq
\int d^4xdz\left \{\sqrt {g}\bigl [-(\partial\phi)^2-m^2\phi^2\bigl ]+
\delta(z-z_0) \sqrt{g_0}L_0(\phi(z))+\delta(z-z_1)\sqrt{g_1}L_1(\phi(z))
\right \}
\label{phiaction}
\eeq
where $L_{0,1}$ are terms localized at the boundaries. Assume that
the dynamics of such boundary terms is such as to fix the values 
$\phi(z_0)=\tilde v_0$ and $\phi(z_1)=\tilde v_1$. In the vacuum the
field $\phi$ will have a bulk profile satisfying
 the 5d Klein-Gordon equation and with $\tilde 
v_0$, $\tilde v_1$ boundary values. The solution is then given by
\beq
\phi=Az^{4+\epsi}+Bz^{-\epsi}
\label{solution}
\eeq
with $\epsi=\sqrt{4+m^2L^2}-2\simeq m^2L^2/4$ for a small mass.
The boundary conditions fix
\beq
A=z_1^{-4-\epsi}\frac{\tilde v_1-\tilde v_0 (z_0/z_1)^{\epsi}}{1-
(z_0/z_1)^{4+2\epsi}}\, ,\quad\quad\quad
B=z_0^\epsi\frac{\tilde v_0-\tilde v_1 (z_0/z_1)^{4+\epsi}}{1-
(z_0/z_1)^{4+2\epsi}}
\label{aandb}
\eeq
and eq.~(\ref{phiaction}) evaluated on the solution yields an effective 
potential for the moduli  $z_0$ and $z_1$
\bea
V(z_0,z_1)
&=&\frac{1}{1-(z_0/z_1)^{4+\epsi}}\left [(4+\epsi)z_1^{-4}\bigl (v_1-v_0
(z_0/z_1)^\epsi\bigr )^2+\epsi z_0^{-4}\bigl (v_0-v_1(z_0/z_1)^{4+\epsi}
\bigr )^2\right ]=\cr &=&z_0^{-4} F(z_0/z_1)
\label{vbare}
\eea
where $v_{0,1}=L^{3/2}\tilde v_{0,1}$ are the boundary vacuum expectation 
values (VEVs) in units of
the $AdS$ curvature. The minimum of $F$, if it exists, determines the size
of the compact dimension. Notice that the value of $F$ at this point,
plays the role of an effective 4d  cosmological constant. Stationarity in 
the conformal factor $z_0$ as well requires a vanishing $F$ at the minimum.
As usual this can be achieved by fine tuning extra 
contributions to the effective
potential. For instance by changing the tensions of the branes with respect
to the RS values $T_0=-T_1=24M^3k$ one gets a correction
\beq
\delta V(z_0,z_1)= z_0^{-4}\delta T_0+z_1^{-4}\delta T_1
\label{modtension}
\eeq
so it is enough to properly choose $\delta T_0$ to get a vanishing effective
cosmological constant at the minimum of $V$. Since we are not interested in
the 4d gravitational dynamics and in the cosmological constant
we will freeze $z_0$ to a constant $=1/\mu_0$ and keep $\mu=1/z_1$ as
our radion field. We are interested in a situation where a 
huge hierarchy $\langle \mu\rangle =\mu_1 \ll \mu_0$ arises. This can be naturally achieved
if $|\epsi|$ is somewhat smaller than 1. In the region $\mu\ll \mu_0$
 we can expand eq.~(\ref{vbare}) as
\beq
V=\epsi v_0^2\mu_0^4 +\left [(4+2\epsi)\mu^4 \bigl(v_1-v_0(\mu/\mu_0)^\epsi
\bigr )^2-\epsi v_1^2\mu^4\right ] +{\cal O}(\mu^8/\mu_0^4)
\label{vren}
\eeq
where in estimating the remainder we have assumed that $|\epsi|\ll 1$.
For $\epsi>0$ the above potential has a minimum around $\mu=\mu_0
(v_1/v_0)^{1/\epsi}$. The hierarchy $\langle \mu\rangle /\mu_0\sim M_W/M_P= 10^{-17}$
can be naturally obtained for fundamental parameters
not much smaller than one (ex. $v_1/v_0\sim 1/10$ and $\epsi\sim 1/20$).
Notice that the hierarchy originates because the relevant part
of the potential is of the form $\mu^4 P(\mu^\epsi)$, i.e. a basically
scale invariant function modulated by a slow evolution through the $\mu^\epsi$
terms. This is very much what happens in the Coleman-Weinberg mechanism 
\cite{coleman},
where a slow RG evolution of the scalar potential parameters
can generate wildly different mass scales.
As we will see below this is not surprising from the CFT point of
view of GW. 

Notice that in $AdS$ one can also consider ``tachyonic'' scalars provided
$m^2>-4k^2$ without introducing instabilities \cite{witten}\footnote{This also
works in a slice of $AdS$ by imposing Dirichlet boundary condition 
$\delta \phi=0$ at the 
Planck brane. This is the case of the GW scalar. On the other hand,
a scalar with $\partial_z \phi=0$ at both boundaries is unstable for
$m_\phi^2<0$. }.
 So in eq.~(\ref{vren})
we can also consider $\epsi<0$. One finds that for 
this case the minimum sits at $\mu=0$.
 However, this does not mean that the $\epsi<0$
cannot be used. We can (should) always imagine that there are extra terms 
coming
from the brane tensions as in eq.~(\ref{modtension}). It is easy to see that
for a range of $\delta T_1$ there is a global minimum at finite $\mu$
also when  $\epsi<0$.

The above calculation of the radion potential neglects the backreaction
of the background metric on the scalar energy momentum density.
The latter is proportional to $\tilde v_{0,1}^2={\cal O}(\tilde v^2)$.
So, based on simple dimensional analysis, we expect our potential to
be the leading result in an expansion in $\tilde v^2/M^3$. One can easily 
estimate the corrections to the radion (and massless graviton) effective
action, by integrating  out at tree level 
the KK modes around the GW approximate solution.
It is easy to see that  lowest order exchange of the massive gravitons yields
a correction to the potential $\delta V\sim \tilde v^4/M^3$, which has
the expected suppression with respect to the leading result,
eq.~(\ref{vbare}). The leading corrections to the radion kinetic term
come instead from the KK excitation of the scalar $\phi$. Notice first of all
that there is no approximate zero mode in this sector, the lowest excitation
has a mass of order $1/z_1=\langle \mu\rangle $, so this procedure makes sense.
Now, since  for  spacetime independent moduli $z_{0,1}$ 
 the scalar action is stationary around eq.~(\ref{solution}), there can 
only be a kinetic
mixing between $\mu$ and the KK modes. By integrating these out
one gets a leading a correction $\sim \tilde v^2 L^3 (\partial \mu)^2$,
 which should be compared to the lowest order result $L_{kin}\sim M^3L^3
(\partial \mu)^2$. In Appendix A we will discuss in some more detail
the radion mass generated by the GW mechanism. We will also comment
on the particular set up \cite{dewolfe} where the backreaction
is automatically included.

In the next Sections, using holography, we will give a purely 4d 
interpretation of the previous results. 
\section{\large The holographic interpretation of the RSI model}

Using the standard rules of $AdS$/CFT, we interpret the
fifth coordinate of $AdS$ as an energy scale.
The region between the two branes 
represents the energy regime $\mu_0\gg E\gg \mu_1$
 where the 4d theory is well approximated by a CFT.
The Planck brane represents the UV cut-off. Its dynamics determines
the boundary conditions for bulk fields: by holography these correspond
to boundary conditions on the coefficients of deformations of the CFT.
Moreover fields that are just localized on the Planck brane 
look like external moduli from the point of view of the CFT: these
fields are very much like a hidden sector coupled to the CFT by $1/\mu_0$
suppressed operators\footnote{This is not true if there are
gauge fields in the bulk \cite{alex,kss,apr} under which the Planck
brane fields are charged. In this case the
coupling between the hidden sector and the visible TeV brane is only
 suppressed by $1/\ln (\mu_0/\mu_1)$.}.
Finally the TeV brane roughly describes the IR limit of the CFT, very much
like the chiral Lagrangian does in QCD.  

The TeV brane abruptly ends $AdS$ space, signalling breakdown of 
conformal invariance in the IR. The first question then is: what
kind of breaking is this? Explicit (soft) or spontaneous?
The first possibility is the same as saying that some  relevant
deformation is turned on in the CFT, eventually generating
a mass gap. This scalar operator would have to be associated through
$AdS$/CFT to some 5d scalar with negative mass${}^2$. The minimal RSI model
(though phenomenologically unacceptable) is perfectly consistent  without
any such scalar. This leaves spontaneous breaking as the only viable 
option. It is also intuitively clear that this is what happens.
The position of the IR brane, which sets the mass scale of the model
is determined by the expectation value of a dynamical radion field. 
All KK masses scale like $\mu_1=\langle \mu\rangle $ while the coupling among
them scales like $1/(c\mu_1)$, with $c\sim(ML)^3$.
In minimal RSI this field is an exact modulus, so it is naturally
interpreted as the Goldstone boson of broken dilatation invariance.
 
We can consider the RSI model as an
 idealized description of a CFT along an exactly
flat direction parameterized by the radion.
Situations in the $AdS$/CFT correspondence where the conformal invariance
is spontaneously broken can be easily found. For example,
the Coulomb branch of N=4 SYM, described by
D3-branes sitting at different points in ten dimensions. 
In this case, the adjoint scalars of
N=4 SYM have a VEV and the radion parameterizes the overall
magnitude of the scalar VEVs. This is a analogy that should be taken
with some care. The RSI model is not supersymmetric and the flat
direction is not necessarily associated with a Coulomb branch or 
a scalar VEV.

We can make several checks of the spontaneous breaking of conformal invariance by adapting the rules of $AdS$/CFT to our case. 
$AdS$/CFT allows to compute Green's functions of composite operators
of the CFT. We can then check that the trace of the stress-energy tensor
is unchanged by the presence of the IR brane.
Moreover, the dilatation current
two-point function has a pole, as dictated by Goldstone's theorem.
In order to 
simplify things we can decouple 4d gravity
by sending the UV brane all the way to the $AdS$ boundary at $z=0$.
We now compute the Weyl anomaly of the dual quantum field theory 
following \cite{hs}. We can restrict our analysis to the UV region.
An explicit breaking of conformal invariance would affect the
trace of the stress-energy 
tensor and would be already visible in the UV.   
The rules of $AdS$/CFT  state that the CFT partition
function in presence of a gravitational background $g_{\mu\nu}^{(0)}(x)$
is given by the classical 5d action $S[g^{(0)}]$ for a solution
\beq
ds^2=\frac{L^2}{z^2}\left (g_{\mu\nu}(x,z)dx^\mu dx^\nu+dz^2\right )
\label{gnframe}
\eeq
of the 5d Einstein's equations satisfying the boundary condition
$\lim _{z\to 0} g_{\mu\nu}(x,z)=g^{(0)}_{\mu\nu}(x)$ plus the requirement that
$g_{\mu\nu}(x,z)$ behaves well 
at the $AdS$ horizon $z\to \infty$.
In our case the latter condition will be replaced by the orbifold Israel
junction conditions at the IR brane. In $AdS$/CFT, Weyl invariance
shows up in the fact that two conformally related boundary metrics
$g^{(0)}_{\mu\nu}(x)$ and $e^{2\sigma(x)}g^{(0)}_{\mu\nu}(x)$ determine 
the same
solution, up to diffeomorphisms. This can be easily seen at the 
infinitesimal level by working around exact $AdS$. Consider the 
coordinate change
\beq
x^\mu={x'}^\mu+f^{\mu}(x')A(z)\quad\quad\quad z=z'-f(x')A_{z'}(z') 
\eeq
with $A_{z'}=dA/dz'$, $f^\mu=\eta^{\mu\nu}\partial_\nu f$ 
and with $A(z')\to -{z'}^2/2$ for $z'\to 0$ so that the boundary
$z=0$ is kept fixed. The  $AdS$ metric of eq.~(\ref{e1}) is changed to
\beq
ds^2=\frac{L^2}{{z'}^2}\left \{(1+2f(x')A_{z}(z')/z')\eta_{\mu\nu}+
2A(z')f_{\mu\nu}(x'))d{x'}^\mu d{x'}^\nu+[1+2f(x')\partial_{z'} (A_{z'}/z')]
d{z'}^2\right \}.
\eeq
Now by choosing exactly $A(z')=-z'^2/2$ the metric is again of the form
in eq.~(\ref{gnframe}) but the corresponding boundary value is now
$g^{(0)}_{\mu\nu}=(1+2f(x'))\eta_{\mu\nu}$. The presence of the IR brane does
not change things much: the only difference is that in the new coordinates
the position of the IR brane is  given by $z'\simeq z_1+f(x)A_z(z_1)$.
However, even though the IR brane is apparently ``bent'', the induced
metric on it is still flat since the bulk geometry has not changed.
 The above argument can be easily extended to infinitesimal 
rescalings around arbitrary $g^{(0)}_{\mu\nu}(x)$. 

By the above argument the partition function $S[g^{(0)}]$ in our case, 
like in standard $AdS$, should be invariant under Weyl rescalings.
Things are however slightly more involved since $S[g^{(0)}]$ is divergent
and must be regulated. The divergences are originated by the region
 $z\to 0$ where the $AdS$ ``volume'' element grows very fast. The regularization
can be done by restricting the integration to the region $z>\epsilon$,
after which the action can be renormalized as $\epsilon\to 0$
 by adding local counterterms covariant in $g^{(0)}$. In usual $AdS$/CFT
it has been shown \cite{hs} that at the end of this procedure
the renormalized  action $S_R[g^{(0)}]$ is Weyl invariant up to an anomaly
local and covariant in $g^{(0)}$.
This is precisely what  one expects for a CFT in a non trivial background.
Now, it is easy to verify that the Weyl anomaly calculation of ref. \cite{hs}
goes through unchanged in the presence of our infrared brane. That
calculation is based on the expansion of the solution $g_{\mu\nu}(x,z)$
around $z=0$
\beq
g_{\mu\nu}=g_{\mu\nu}^{(0)}+z^2g_{\mu\nu}^{(2)}+z^4g_{\mu\nu}^{(4)}
+z^4\ln z h_{\mu\nu}^{(4)}+\cdots.
\eeq
The point is that the anomaly depends only on $g^{(2)}$ and 
${\mbox Tr}g^{(0) -1}g^{(4)}$, which algebraically depend
on $g^{(0)}$, and the boundary conditions at the IR brane (which start 
to matter at order $z^4$) do not affect these terms.
We conclude that the RSI model in the limit
where 4d gravity decouples is conformal. 

The conformal symmetry is however non-linearly
realized in RSI. This can formally be seen by the fact that under the 
conformal transformation $x\to \lambda x$, $z \to \lambda z$, 
the $AdS$ isometry, the position of the IR brane is changed $z_1\to \lambda z_1$.
The physics is however unchanged, so that $z_1$ parameterizes a manifold of 
equivalent vacua. The associated Goldstone boson is the radion $\mu$.
This can more directly be seen by considering the effective Lagrangian
for radion and 4d gravity calculated in refs. \cite{csaki0,goldwise}
\beq
\L={\sqrt g}M^3L^3\left \{2(\mu_{0}^2-\mu^2)R(g)-12(\partial \mu)^2\right \}.
\eeq
When $\mu_0\to \infty$ 4d gravity decouples and $g$ is just a
background probing our CFT (consistently, the induced metric on
the Planck brane is exactly $g$).  For this purpose notice 
that the $\mu$ dependent terms are formally Weyl invariant
\beq
{\sqrt {g}}M^3L^3\left \{-2\mu^2 R(g)-12(\partial \mu)^2\right \}=
-{\sqrt g}M^3L^3\mu^4R(\mu^2 g).
\label{weylradion}
\eeq
Moreover $\mu$  couples to the TeV brane only in the Weyl invariant 
combination $\mu^2 g_{\mu\nu}$.
By integrating out the radion at tree level eq.~(\ref{weylradion})
gives a contribution $S_{rad}[g]$ to the source action
\beq
S_{rad}[g]=M^3L^3\int {\sqrt g}d^4 x \left (
-2\mu_1^2 R(g)- \frac{\mu_1^2}{3}R(g)
\Dal^{-1} R(g)\cdots\right ).
\label{eff}
\eeq
$S_{rad}$  is invariant under Weyl transformations $g\to e^{2\sigma(x)} g$ with
$\sigma(x)$ vanishing fast enough at infinity,
but not under strictly rigid rescaling $g \to \lambda g$. The rigid dilatations
cannot be smoothly obtained from the local ones because of surface terms
in the variation of $S_{rad}$, a signal of spontaneous symmetry breaking.
By varying $S_{rad}$ twice with respect to $g_{\mu\nu}$  we obtain the 
radion contribution to the two-point function of $T_{\mu\nu}$.
Eq.~(\ref{eff}), being Weyl invariant,
only contributes to the transverse-traceless part of the
correlator:
\beq
\langle  T_{\mu\nu} (x) T_{\rho\sigma}(0)\rangle=-{2\over 3}(ML)^3\mu_1^2
{\prod}^{(2)}_{\mu\nu\rho\sigma}{1\over \Dal}\delta^4(x)
\label{pippo}
\eeq
where $\pi_{\mu\nu}=\partial_\mu\partial_\nu -\eta_{\mu\nu}\partial^2 $
is a projector over traceless tensors, and 
${\prod}^{(2)}_{\mu\nu\rho\sigma}=2\pi_{\mu\nu}\pi_{\rho\sigma} -3
(\pi_{\mu\rho}\pi_{\nu\sigma}+\pi_{\mu\sigma}\pi_{\nu\rho})$. In analogy with
current algebra, the Goldstone boson creation amplitude is 
\beq
\langle 0|T_{\mu\nu}|rad\rangle =\sqrt{\frac{8}{3}}(ML)^{3/2}\mu_1 p_\mu p_\nu
\eeq
and $\sqrt{8/3}(ML)^{3/2}\mu_1$ is the radion decay constant.
The same Goldstone pole appears in the correlator 
$\langle  S_\mu(x) S_\nu(-x)\rangle$ of the dilatation
current $S_\mu=x^\nu T_{\mu\nu}(x)$.

The same pole should appear in the energy momentum correlator
computed using the rules of $AdS$/CFT.
The two point function of a conserved stress-energy tensor has the 
general form \cite{a3} 
\begin{equation}
\langle  T_{\mu\nu} (x) T_{\rho\sigma}(0)\rangle  = -{\frac{1}{48\pi^4}}{\prod}%
^{(2)}_{\mu\nu\rho\sigma} \left[{\frac{c(x)}{x^4}}\right] +
\pi_{\mu\nu}\pi_{\rho\sigma}\left[{\frac{f(x)}{x^4}}\right],
\label{duepunti}
\end{equation}
For a conformal theory, $f=0$ and $c(x)$ is constant and equal to
the central charge $c=4\pi^2(ML)^3$ 
of the theory. In a spontaneously broken 
theory, $f$ still vanishes but $c(x)$ is non-trivial.
>From eq.~(\ref{pippo}), we expect that the Fourier 
transform of $c(x)/x^4$ has a pole $1/p^2$ for $p\rightarrow 0$. 
Let us compute this quantity using the rules of $AdS$/CFT. 
The computation of the boundary effective action is essentially
like in ref. \cite{witten}. The horizon region has been replaced 
with the TeV
brane, and the field behavior at the horizon replaced by the orbifold
conditions at the TeV brane.
The transverse-traceless part of the two-point function of $T_{\mu\nu}$
can be read from the two-point function of a minimally coupled
scalar field, which is explicitly computed in Appendix B.
The results is 
\beq
{\cal F}(p)=\int d^4p e^{ipx}{c(x)\over x^4}=-2\pi^2c
\left (\log {p\over 2} -{K_1(p/\mu)\over I_1(p/\mu)}\right )
\eeq
which has indeed a pole ${\cal F}(k)\sim {16\pi^4\mu^2(ML)^3\over p^2}$
with the right coefficient
\footnote{The factor
2 mismatch is trivially due to the doubling of the RS action on the orbifold with
 respect to the standard $AdS$/CFT computation where only one
slice of $AdS$ is present.}.

There is still an important point to be addressed.
By the above discussion, corresponding to a radion 
value $\langle \mu\rangle =\mu_1$ there should be some CFT operators 
$O_i$ of dimensions
$d_i$ getting VEVs $\langle O_i\rangle =c_i\mu_1^{d_i}$, with $c_i$ constants.
(This scaling in $\mu$ is due to the absence of explicit conformal breaking
sources or, better, to conformal covariance.). The large euclidean momentum
behavior of correlators from the point of view of the OPE gives
information on the dimensions of the $O_i$. For the two point function 
at momentum $p$  of an operator of dimension $d$ we expect
\beq
\langle O_d O_d\rangle _{p^2}=\frac{b_0}{p^{4-2d}}+\sum_i b_i \frac{\langle O_i\rangle }{p^{4-2d+d_i}}.
\label{ope}
\eeq 
Such correlators can be calculated using the holographic prescription.
The example for a dimension four operator is discussed in Appendix B.
The general result \cite{apr} is that at euclidean 
$p\gg\mu$ the deviation from conformality is suppressed
like $\exp (-p/\mu)$ rather than being just power suppressed as naively
expected from eq.~(\ref{ope}).
The 5d reason behind  this wild suppression
is that the local geometry between the two branes is {\it exactly}
$AdS$5. The natural 4d interpretation of this result is that 
the operator that spontaneously breaks conformal invariance has 
formally infinite dimension. 
In $AdS$/CFT small curvature 
on the $AdS$ side corresponds to large $N$ and large 't Hooft parameter $g N$ 
in the CFT. So it is conceivable that the operators involved in
symmetry breaking have inverse dimensions $1/d$ which vanish
at lowest order in the large $N$ and large $gN$ expansion
\footnote{All the string states, which certainly populate the $AdS$ background,
 correspond to operators with large dimension $\sim\sqrt{gN}$
\cite{witten}. The operators $O_i$, corresponding 
to a set of supergravity
modes with large mass, would certainly mix with string states
in the IR. As a toy model, we could consider
a distribution of D3-branes in type IIB. A quasi-spherical
distribution with only high multipoli momenta gives VEV only
to the operators $O_k=Tr(\phi_{\{ i_1}...\phi_{i_k\}})-{\mbox traces}$, 
$k\ge k_0$ and we can take $k_0$ arbitrarily large. 
A completely spherical distribution has all the multipoli
VEVs $\langle O_k\rangle =0$ but has $tr\phi_i\phi_i\ne 0$. It is well known
that the trace $tr\phi_i\phi_i$ is the prototype of a stringy mode in
the $AdS$/CFT correspondence for N=4 SYM.
It would be interesting to find an explicit type IIB model
which mimics all the relevant features of the RSI model.}.
In the $AdS$ side finding 
the dimensions of these operators requires a smoothing of the jump
singularity at IR brane by lifting the RS model to 10 dimensions and
by accounting for string modes.

\section{\large GW radius stabilization: holographic interpretation}

We saw that conformal breaking in the vacuum is due to
the vevs $\langle O_i\rangle =c_i\mu^{d_i}$ of one or more operators with
a ``large'' dimension. In general, corresponding to this field
configuration, there will be an effective potential. By conformal 
invariance this must be of the form $V_{eff}=a\mu^4$, with $a$ constant. 
Of course 
in minimal RSI, $a= 0$ and all $\mu$'s have the same energy. 
However changing the TeV brane tension, 
like in eq.~(\ref{modtension}),
corresponds to having $a=\delta T_1$. Then for $a>0$ the TeV brane
is pushed to the horizon and conformal invariance is unbroken 
in the vacuum, while for $a<0$ the system is destabilized 
(the TeV brane falls towards 
the Planck one). So, only for $a=0$ is conformal invariance truly
spontaneously broken with the radion being the corresponding Golstone
boson. 

By the above discussion, in order to get both $\langle \mu\rangle \not = 0$ 
and no Goldstone
mode, it is necessary to introduce some explicit source of conformal breaking.
The simplest thing to do is to add a  perturbation
\beq
\delta L=\lambda {\cal O}
\eeq
where ${\cal O}$ is an operator of dimension $[{\cal O}]=4+\epsi$.
For $|\epsi|\ll 1 $ we may call the perturbation  {\it almost marginal}.
The coupling evolves with the renormalization scale $Q$ as 
$\lambda(Q)=\lambda(\mu_0) (Q/\mu_0)^\epsi$. By RG invariance
the effective potential will now have the form
\bea
V_{eff}&=&\mu^4 P(\lambda(\mu))=a\mu^4+\mu^4\sum_{n=1}a_n\lambda^n(\mu)=\cr
&=&\mu^4\left (a+a_1\lambda(\mu_0)(\mu/\mu_0)^\epsi+a_2\lambda(\mu_0)^2
(\mu/\mu_0)^{2\epsi}+a_3\lambda(\mu_0)^3(\mu/\mu_0)^{3\epsi}+\cdots\right ).
\label{colwein}
\eea
Notice that the finite part of eq.~(\ref{vren}) is just a special case 
of this where $\lambda(\mu_0)= v_0$. The discussion of the previous section
on the possibility of generating minima with $\mu/\mu_0\ll 1$ applies.
Of course we would need to be able to calculate the coefficients $a_i$
in the CFT in order to say something on the vacuum.

What we have outlined in the previous paragraph is just the
4d picture of the GW mechanism. In the $AdS$/CFT dictionary
a bulk scalar $\phi$ with mass $m^2$ corresponds to a scalar 
operator ${\cal O}$
of dimension $d={\sqrt {4+m^2/k^2}}+2=4+\epsi$. The 
two independent solutions $z^{4+\epsi}$ and $z^{-\epsi}$ of $\Dal \phi=0$
in the $AdS$ background  are respectively associated to the
VEV $\langle \O\rangle $ and to the source $\lambda$. More precisely, given
 the profile
\beq
\phi=Az^{4+\epsi} + Bz^{-\epsi}
\eeq
$AdS$/CFT relates \cite{witten,klebanov}
\beq
A=\frac{\langle \O\rangle }{4+2 \epsi}\quad\quad B=\lim_{\mu_0\to \infty}
\mu_0^{-\epsi}\lambda(\mu_0).
\label{dictionary}
\eeq
Confronting with eq.~(\ref{aandb}), we basically have that 
$\tilde v_0=\lambda(\mu_0)$
is the UV deformation parameter. Moreover by taking the limit $\tilde v_0=0$
we have $\langle \O\rangle =(4+2\epsi)\mu^{4+\epsi}\tilde v_1$. So the two vevs $v_0$
and $v_1$ are respectively associated to explicit and spontaneous breaking of
conformal invariance.

Eqs.~(\ref{dictionary}) can be checked directly using
the holographic potential in eqs. (\ref{vbare},\ref{vren}). In order to avoid
spurious cut-off effects one should take the
limit $z_0\to 0$ with the scaling $v_0=\bar v_0 z_0^{-\epsi}$
and also renormalize eq.~(\ref{vbare}) by subtracting the $\mu$ independent
$1/z_0^4$ divergent term. One gets
\beq
V_{ren}=\left [(4+2\epsi)\mu^4 \bigl(v_1-v_0(\mu/\mu_0)^\epsi
\bigr )^2-\epsi v_1^2\mu^4\right ]
\label{truev}
\eeq
from which
\beq
\langle \O\rangle =-\mu_0^\epsi\frac{\partial}{\partial \lambda(\mu_0)} V_{eff}
\equiv -z_0^{-\epsi}\frac{\partial}{\partial v_0} V_{ren}=
2(4+2\epsilon)(v_1-\bar v_0\mu^\epsi)
\eeq
which is consistent with eqs.~(\ref{aandb},\ref{dictionary})\footnote{Once
again, the factor
2 mismatch is due to the doubling of the RS action on the orbifold,
with respect to ref. \cite{klebanov} where there is 
just one branch of $AdS$.}.
 
Finally one should not be puzzled by $V_{ren}$ being only quadratic 
in the perturbation. This is because the
 GW scalar is a free field and correspondingly
the connected Green's functions with more than two $\O$ legs vanish 
\cite{witten}. By considering a self-interacting $\phi$ we would
get an effective potential of the general form (\ref{colwein}).
In Appendix A we further discuss the radion potential and mass in
the general case.

By considering its 4d picture we can better understand in which sense the
RS model solves the hierarchy problem. Conformal symmetry as opposed to
supersymmetry was invoked by Frampton and Vafa
\cite{framptonvafa} as a principle to
cure the destabilizing quadratic divergences of the SM. However, 
as we will now discuss,
it does not work as well. It is certainly  acceptable
to elevate the conformal group to a fundamental symmetry. Then this
symmetry has to be valid at all energy scales somewhat larger than
the weak scale. However this does not help the hierarchy problem,
which is one of separation of mass scales, {\it i.e.} the separation of
the Fermi scale from the Planck or GUT scales. The presence of these 
other scales implies that the theory can only be conformal
in a limited energy regime. This is to say that the conformal symmetry
is not fundamental but dynamical, and we cannot use it as a principle
to discard unwanted parameter choices. In general these perturbations
will be there and in order for the model to be interesting they
should not badly affect the hierarchy. 
Then the question  whether a CFT can solve the hierarchy problem
depends on the classification of its relevant deformations.
If the theory admits deformations of dimension 2 or 3, which cannot
be discarded by independent symmetry considerations, then the electroweak
 hierarchy
is badly unstable. This is truly the case of the SM, which being weakly
coupled is almost
a CFT, and which admits a dimension 2 Higgs mass deformation.
As shown in ref. \cite{csabajohn} this is also what happens in the
model of ref. \cite{framptonvafa}.
The RS-GW model instead is dual by construction  to a CFT where 
the most relevant
deformation is almost marginal, {\it i.e.}  $|d-4|=|\epsi|\ll 1$.
Such a deformation, see eq.~(\ref{colwein}), determines a large hierarchy 
of scales, essentially like in the Coleman-Weinberg mechanism.
Of course if the GW field had a negative mass such that, say $\epsi=-2$,
the CFT would have the analogue of the Higgs mass problem in the SM.
The minimum of eq.~(\ref{colwein}) would generically be at the cut off scale:
$\langle \mu\rangle \sim \mu_0$. The RS-GW model is a (maybe ad hoc)
construction of a CFT without strongly relevant deformations. However
it gives a fairly predictive set up, at least when the $AdS$ geometry
is not too curved, {\it i.e.} when the CFT has both large N and large 
't Hooft coupling. So it is worth further investigations.

\section{\large Shining and Natural Flavor Conservation}

Any attempt at solving the hierarchy problem by invoking new physics
just above the weak scale runs the risk of spoiling the good
features of the SM in the matter Flavor, Baryon and Lepton quantum numbers.
In the SM the only ``relevant'' operators that violate flavor are the
Yukawa couplings. This property
goes under the name of Natural Flavor Conservation (NFC).
 All dangerous effects are properly suppressed
by combinations of quark masses and CKM mixing angles and the resulting
phenomenology agrees with the data. Similarly, baryon and lepton numbers
are violated at lowest order respectively by operators of dimension
6 and 5. Provided the SM cut-off is large these properties nicely explain
why we have not yet detected proton decay and why the neutrini are so light.

In the flavor sector probably the most natural assumption is that NFC
keeps being valid beyond the SM. For example this possibility
is realized by models with gauge mediated supersymmetry breaking\cite{dn}.
Composite technicolor models \cite{comptec} also aim at this  principle.
We can do the same thing in the RSI/CFT model and see 
what the consequences are. By working in the CFT picture, NFC 
corresponds to the requirements
\begin{enumerate}
\item There is an approximate global 
$G_F=SU(3)^5$ flavor symmetry. This symmetry
acts in the usual way on the low energy degrees of freedom, quarks and leptons.
(By assuming $G_F=SU(3)^5\times U(1)_B\times U(1)_L$ we could
also take care of baryon and lepton number, as discussed below.)

\item The CFT scalar operators with non trivial $G_F$ quantum numbers
are all strongly irrelevant, say  of dimension $>5$. The only
exception is represented by 3 multiplets of almost marginal 
operators $\O_u$, $\O_d$, $\O_e$,
with the quantum numbers of the SM Yukawa couplings ({\it i.e.} in 
$(\bar 3,\bar 3)$-type representations of the flavor group).
\end{enumerate}
By these requirements, the CFT can be defined also in a limit where
flavor is unbroken. Then by turning on at the cut-off the most general sources
of flavor violation one is guaranteed that at low energy the only
effects that are not strongly power suppressed are due to the sources 
$Y_u$, $Y_d$ and $Y_e$ of the $\O_u$, $\O_d$, $\O_e$. 
These are almost marginal and can survive
(or even grow a little) over several decades of RG evolution.

By applying the $AdS$/CFT dictionary points 1) and 2) translate
into the following requirements on the RSI model.
\begin{enumerate}
\item $G_F$ is gauged in the 5d bulk. This gauge symmetry corresponds
to having the same symmetry, but global, in the CFT:
the 5d flavor gauge bosons are dual to the global  4d flavor currents.
\item There are 5d scalars $\phi_u$, $\phi_d$ and $\phi_e$ transforming
as the corresponding Yukawas. They have bulk masses $m^2_{u,d,e}$
such that $\epsi_{u,d,e}={\sqrt{4+m_{u,d,e}^2L^2}}-2$ is small
in absolute value. Indeed it is enough that $|\epsi_{u,d,e}|$
be smaller than the corresponding $|\epsi|$ of the GW field. 
This makes sure that Yukawas do not get too depressed by running
down from the Planck scale, or that they do not destabilize the hierarchy.
\item 
$G_F$ is spontaneously broken to the identity by scalar fields
living on the Planck brane {\it only}.
This breaking generates at the Planck brane sources for the $\phi$'s 
so that Yukawa couplings are generated on the TeV brane by shining 
\cite{nimasavas}
through the 5d bulk. The origin of flavor breaking cannot be in the bulk
or on the TeV brane, cause that would mean that the inner CFT dynamics
breaks flavor. Instead we want flavor to be broken by external
sources, so it should happen on the Planck brane only.
\end{enumerate}

We should say that we are not trying to explain the structure of fermion
masses and their mixing angles. In the above description the explanation
for that lies within the Planck dynamics. For instance one could imagine 
the Planck brane to consist of several branes all separated by some distance
and then apply the ideas of refs. \cite{nimasavas,ahsw}. However, for the 
general 
predictions we will discuss, all we need to know is that at the Planck brane
(or at some $z\ll z_1$) the b.c. on the scalars already reproduce the
SM Yukawa structure
\beq
\phi_{u}(z_0)=a_uk^{3/2} Y_u\quad\quad \phi_{d}(z_0)=a_dk^{3/2} Y_d\quad\quad
\phi_{e}(z_0)=a_ek^{3/2} Y_e
\eeq
with $a$'s of order 1. Then the scalars in the bulk behave at lowest
order in $Y$ like (ex. $\phi_u$)
\beq
\phi_u =\left (Az^{4+\epsi_u}+Bz^{-\epsi_u}\right ) Y_u
\eeq
where $A,B$ are numerical coefficients fixed by the boundary conditions.
For instance if the condition on the TeV brane is just $\partial_z \phi=0$
we have that $A$ is negligible and that $B\simeq a_uk^{3/2}z_0^{\epsi_u}$.
In this case $\phi_u\propto Y_u$ is essentially constant through 
the bulk. (We will further discuss the TeV brane b.c. in the next section).

The reason why this set up realizes NFC is that the only sources  
of flavor violation that are active close to the TeV brane are the scalar
profiles, i.e. the SM Yukawa couplings. Consider for instance the masses
of the $SU(3)^5$ KK vectors. First of all since the group is completely
Higgsed there are no zero modes. Moreover the modes with mass $M\ll \mu_0$
are localized close to the TeV brane, so that they are totally insensitive
to the original sources of Flavor breaking on the Planck brane.
These masses are only (mildly) sensitive to the 
$\phi_{u,d,e}\propto Y_{u,d,e}$ profiles. For small Yukawa the masses are 
given by
\beq
m_n=\mu_1 \left (x_n+O(Y^2)\right)
\eeq
where $x_n\simeq (n-1/4)\pi$ are the roots of $J_0(x)=0$. Notice that there
is no massless vector even for $Y\to 0$ \cite{alex0}, as long as $G_F$
is totally broken on the Planck brane. For instance
vector exchange will induce flavor symmetric effective operators
up to corrections quadratic in the Yukawa couplings and respecting the 
$SU(3)^5$ selection rules. Consistency with the low energy phenomenology
is easily met for vector masses of order a TeV \cite{ahsw}\footnote{Indeed
the severest constraint comes from electric dipole moments. These are 
however easily satisfied if one assumes that CP violation is mediated to
the SM only through the $\phi_{u,d,e}$ profiles.}.
Similar conclusions are reached for the KK couplings to the SM fermions.
Up to small Yukawa corrections these are $SU(3)^5$ symmetric and given by
\beq
g_i^2=\frac{\bar g_i^2}{L}\quad \quad {i=1,\dots,5}
\eeq
where $\bar g$ are the 5d couplings. We conclude that NFC leads in the RS
scenario to a reach set of predictions which are in principle testable
at future colliders.

Notice that each factor $SU(3)_i$ in $SU(3)^5$ has a cubic anomaly $a_i$
from the SM fermions localized on the TeV brane. This localized anomaly can be
taken care of by the proper Chern-Simons (CS) term in the bulk
\beq
L_{CS}=\frac{a_i}{16\pi^2}\int \left (A\wedge d A\wedge dA+\cdots\right ).
\eeq
Then there should  also be a $-a_i$ anomaly due to fermions on the Planck
brane. These fermions are however of no consequence: they have
mass $O(\mu_0)$ as $G_F$ is broken  on the UV brane, and moreover they couple
very weakly \cite{alex0,alex} to the light KK vector bosons.

If we want to explain baryon and lepton number conservation
we should also gauge them. However both $U(1)_B$ and
$U(1)_L$ have mixed SM anomalies like for instance $U(1)_B SU(2)_W^2$
which cannot be eliminated by a CS term. This is because the SM gauge
group is localized on the brane. For $B$ and $L$ we are forced to
add extra matter charged under $SU(2)_W\times U(1)_Y$
on the TeV brane to cancel the mixed anomalies \cite{add}.
The $B^3$, $BL^2$, $B^2L$ and $L^3$ can still be cured by the CS term. 
With these proviso we may then imagine that $B$ and $L$
are broken on the Planck brane. Now, to be consistent with proton
decay data, we will have to assume that the lightest bulk scalar
with $B=1$ and $L=-1$  has a mass $m^2L^2>12$. On the other hand
in order to naturally obtain the neutrino mass observed at SuperKamiokande
we need a bulk scalar with $B=0$, $L=2$ and mass 
$m^2L^2\sim 4\div 5$.

\section{\large Radius Stabilization by Electroweak Breaking}

In this section we want to consider some variations over
the minimal GW set up. So far we have assumed that the boundary
potentials $V_0(\phi)$ and $V_1(\phi)$ select some fixed values
$\tilde v_0$ and $\tilde v_1$ for $\phi$. This means
that we considered the limit where
$a_2,b_2\to\infty$ in the expansion
\bea
V_0&=&\frac{a_2}{2}(\phi-\tilde v_0)^2+\frac{a_3}{3}(\phi-\tilde v_0)^3+\cdots
\cr
V_1&=&\frac{b_2}{2}(\phi-\tilde v_1)^2+\frac{b_3}{3}(\phi-\tilde v_1)^3+\cdots.
\label{boundaryv}
\eea
As discussed already in ref. \cite{gw} (see also \cite{dewolfe})
the GW mechanism can work also in the more general case. 
As a different, and interesting, parameter choice let us take 
$b_2$ finite and $\tilde v_1=0$. The latter choice would be
the right one for the flavor fields of the previous section, as
we assumed that flavor is not broken by the CFT inner dynamics.
To simplify things we also assume a small source in order to neglect
cubic and higher terms in the $\phi$ action. The boundary
condition at the TeV brane is 
\beq
-2z_1\partial_{z}\phi=L\partial_\phi V_1(\phi)\simeq b_2 L \phi
\eeq
while at the Planck brane it is still $\phi(z_0)=\tilde v_0$. The solution 
is then given by
\beq
\phi(z)=\tilde v_0 (z_0/z)^\epsi \frac{1-(z/z_1)^{4+2\epsi}\eta }{1-
(z_0/z_1)^{4+2\epsi}\eta}\quad\quad\quad \eta=\frac{b_2L-2\epsi}{8+2\epsi+b_2L}
\eeq
and the (renormalized) effective potential
\beq
V=\eta(4+2\epsi) v_0^2\mu_0^{-2\epsi}\mu^{4+2\epsi}.
\label{neumann}
\eeq
Corresponding to $\tilde v_1=0$ the VEV $\langle \O\rangle \propto  
A=
-\eta v_0\mu_0^{-\epsilon}
\mu^{4+\epsi}$ vanishes with the source $v_0=\lambda(\mu_0)$ and the
potential has no piece linear in $\lambda(\mu)$. (eq.~(\ref{neumann})
agrees with eq.~(\ref{vren}) in the limit $v_1=0$, $b_2\to \infty$.)

Now, it is easy to check that by even taking finite $a_2$ in $V_0$ 
the $\mu$ dependence of the renormalized potential is the same.
Only the coefficient in front is affected. This just corresponds to a
redefinition of the source
\beq
v_0\to f(v_0)=d_1 v_0+d_2 v_0^2+\dots
\eeq
in eq.~(\ref{neumann}). Again, by modifying physics on the
Planck brane we simply modify the definitions of our CFT deformation
parameters, but the infrared behavior ($\mu$ dependence) remains
the same. Without loss of generality we can simply fix the $\phi(z_0)$. 

Eq.~(\ref{neumann}) by the addition of  a tension term $\delta V=\beta \mu^4$
can stabilize $\mu$. For $\epsi>0$ one needs $\eta>0$ and $\beta<0$,
for $\epsi<0$ the reverse $\eta<0$ and $\beta >0$. Notice  that
$\eta<0$ implies $b_2<0$, which does not lead to an
instability as long as $b_2L\lsim 1$.

In the RS model both the hierarchy
and NFC require bulk scalars with a mass $|m^2|\ll 1/L^2$.
So it is natural to assume that the role of the  GW field is 
played by the $\phi_{u,d,e}$ themselves. 
 Then the radion potential will be mainly determined by
the field $\phi_t$ corresponding to the top quark Yukawa coupling. This
field has the ``largest'' profile. By flavor conservation on the TeV
brane the potential $V_1$ in eq. \ref{boundaryv} is a function of 
$\phi_t\phi_t^\dagger$,
so it corresponds to $v_1=0$. At lowest order it gives a radion potential
of the form in eq.~(\ref{neumann}). As we said, stability can be achieved
by balancing this term with $\beta\mu^4$. However with the SM living
on the brane there is also another, more interesting, option.
The Higgs doublet $H$ could be involved in the stabilization dynamics.
In general the coefficients $b_i$ will depend on the bilinear $HH^\dagger$,
so that at quartic order in $H$ the effective potential
for Higgs and radion will be written as
\beq
V=y_1(\mu) \mu^2 HH^\dagger +y_2 (\mu) (HH^\dagger )^2 +y_3(\mu)\mu^4.
\eeq
The dominant $\mu$ dependence of the $y_i$'s is through the top flavon source
$L^{3/2} \phi_t(z_0)= v_0$
\beq
y_1=({y_1})_0+({y_1})_1 v_0^2 (\mu/\mu_0)^{2\epsi}+\cdots.
\label{runmass}
\eeq
For instance one could consider a parameter choice where the $y_i$ 
start positive at $\mu_0$ but $y_1$ goes negative at lower $\mu$.  
By truncating
eq.~(\ref{runmass}) to the first two terms, the choice
$({y_1})_0>0$, $({y_1})_1<0$ $\epsi<0$ achieves the goal. 
For negative $y_1$ one can minimize
in the Higgs field and find an effective potential for $\mu$
\beq
V_{eff}=\left (y_3(\mu)-\frac{y_1(\mu)^2}{4 y_2(\mu)}\right ) \mu^4=
y_{eff}(\mu)\mu^4.
\label{effmu}
\eeq
Then $\langle \mu\rangle $ will roughly correspond to the point (if it exists) where
$y_{eff}$ crosses zero. Notice that a stable minimum now exists
even for $y_3\equiv 0$.
A similar interplay between a dilaton type field
$\mu$ and the Higgs was already suggested in no-scale inspired supersymmetric
models \cite{Antoniadis:1993fh,kz,bs,Antoniadis:1998zg,Antoniadis:1999sd}
where the soft terms are proportional to a modulus.
In particular, in the ``phenomenological''   models in refs. \cite{kz,bs}
 the $\mu$ dependence of the effective potential
couplings  is just determined by RG evolution  in the MSSM. This makes them 
 more predictive than the RS model where we, instead, have to make  
assumptions about  the behavior 
of the $y$'s. However in refs. \cite{kz,bs} a much stronger assumption
on the absence of quadratic divergences in the modulus potential had
to be made. Here this result simply follows from the properties
of the deformations of the CFT, of which $\mu$ is part\footnote{ The models
\cite{Antoniadis:1993fh, Antoniadis:1998zg,Antoniadis:1999sd} are truly 5 dimensional
above the weak scale: this fact and supersymmetry insure that the modulus is not
badly destabilized. In order to mantain a hierarchy between the TeV and Planck scales,
as it happens in the RS model, the 5d theory itself should flow to a fixed point
above a TeV \cite{Antoniadis:1999sd}. It would be interesting to find explict 
examples where this happens and for which the modulus dynamics remains calculable.}.

Now, having the Higgs to play a role in radius stabilization sounds
nice, but leads to no predictions unless further assumptions are made.
One reasonable step then is to consider the limit where $\mu$ is weakly 
coupled. In the $AdS$ picture $\mu$ is part of the gravitational field
and its kinetic term is 
\beq
L_{kin}=-12(ML)^3 (\partial \mu)^2=\frac{N^2}{2}(\partial \mu)^2.
\eeq
We want to consider $N$ large with the $y$'s fixed for which the radion
decay constant $\Lambda_\mu=N\langle \mu\rangle $ is somewhat 
larger that the weak
scale, and the Higgs-radion mixing suppressed. At the minimum of 
eq.~(\ref{effmu}) we find (the dots represent $\mu d/d\mu$ derivatives)
\beq
m_\mu^2\Lambda_\mu^2=(-16y_{eff}+{\ddot y}_{eff})\mu^4=
2 m_h^2 v_F^2 \left ( 1-\frac{4 y_2 y_3}{y_1^2}\right )(1+O(\epsi^2)/y_{eff})
\leq 2 m_h^2 v_F^2
\eeq
where $v_F={\sqrt 2}\langle H\rangle =246$ GeV. The inequality is obtained neglecting 
terms that are generically $O(\epsi)$ and by considering that at the
minimum $y_{eff}<0$ with $y_3>0$  ($0<1-4y_2y_3/y_1^2<1$).
In the limit we considered $m_\mu\sim v_F^2/\Lambda_\mu$, like any modulus.
For $\Lambda_\mu$ of order a few TeV we could reasonably expect a radion
with a mass of a few GeV or even less. It may be worth to devote
more attention to this region of parameter space, which has been 
so far neglected. 

\section{\large Conclusions}
In this paper we have discussed the four dimensional interpretation
of the compact Randall-Sundrum model. Basically we have shown that
the system of Kaluza-Klein states plus the massless radion corresponds
to 4d gravity coupled to a field theory with a spontaneously broken
conformal invariance. This conclusion is easily reached by taking the
limit where the Planck brane is at the boundary of $AdS$, in which 4d gravity 
decouples. In this limit, the reaction of the system to an external
4d metric $\bar g_{\mu\nu}$ can be studied by the standard $AdS$/CFT technique.
The generating functional $S[\bar g]$ so obtained is invariant (up to the
anomaly) under Weyl transformations regardless of the presence of the IR brane,
indicating that the theory satisfies the Ward identities for
conformal invariance. However the presence of the IR brane does not allow
to define conformal transformations globally: a mass gap is generated and
the radion is the Goldstone boson of dilatations.
This result is valid at tree level in the 5d theory. It should however  
remain true by including quantum corrections in an effective field
theory approach (like the one used for the pion Lagrangian) regardless
of whether there is or there isn't a fundamental string theory
description of the model. This is just because of the spacetime symmetry 
properties of $AdS$ space \cite{witten}.
On the other hand, a string theory description would allow a complete
description of the CFT, including operators of arbitrarily high dimension
($\gsim ML$)
and spin. Correspondingly it would allow to study interactions in the
energy regime $E\sim (ML)\mu_1$ where the effective Lagrangian 
approach breaks down.  

The Golberger-Wise stabilization mechanism corresponds to breaking 
conformal invariance explicitly by turning on a coupling 
$\lambda_{\scriptscriptstyle GW}$ of 
dimension
$-\epsilon$. This leads to a non trivial effective potential 
for the radion $V(\mu)=\mu^4P(\lambda_{\scriptscriptstyle GW}
\mu^\epsilon)$, which generically has a minimum at
\beq
m_{weak}=<\mu>\sim \lambda_{\scriptscriptstyle GW}^{-{1}/{\epsilon}}.
\eeq
For $|\epsilon|\ll 1$, a natural hierarchy is generated. 
As $\lambda_{\scriptscriptstyle GW}$ is an essentially dimensionless parameter, one
could say that the hierarchy is generated by dimensional transmutation,
like in the Coleman-Weinberg mechanism or in technicolor. But unlike technicolor the 
coupling $\lambda_{\scriptscriptstyle GW}$ responsible for the hierarchy can
be either strong or weak at the TeV scale (see Appendix A). Which case is
realized is a matter of parameter choices in the gravitational picture. In the
case of a weak $\lambda_{\scriptscriptstyle GW}$ there is a light 
scalar resonance, radion, that couples to the trace of the energy momentum tensor.
It is amusing that Natural Flavor Conservation suggests
that the SM Yukawa couplings are almost marginal. So it is possible
that the role of $\lambda_{\scriptscriptstyle GW}$ in the RS model  is 
played by the top Yukawa coupling.

\paragraph{Acknowledgements}

We would like to thank Nima Arkani-Hamed  and Massimo Porrati for many
important discussions and comments. We would also like to thank 
Csaba Cs\`aki, John March-Russell,
Luigi Pilo, Alex Pomarol and Alessandro Strumia for stimulating conversations.
R.R. thanks the hospitality of the TH division at CERN. This work 
is partially supported by the EC under TMR contract HPRN-CT-2000-00148
(R.R.). 

\appendix
\section{\large The radion mass}
For a generic potential of GW type
$V(\mu)=\mu^4 P(\mu^\epsilon)$ the radion mass is given by
\beq
m_{rad}^2=\frac{1}{24 M^3L^3} \mu_1^{2+\epsilon}\left [
(4\epsilon +\epsilon ^2) P'(\mu_1^\epsilon)+\epsilon^2\mu_1^\epsilon 
P''(\mu_1^\epsilon)\right ]
\label{mass}
\eeq
where we used the stationarity condition $4P(\mu_1^\epsilon)+\epsilon
\mu_1^\epsilon P'(\mu_1^\epsilon)=0$. For small $\epsilon$ the 
minimum is roughly at the place where $P$ vanishes. Generically
the above equation implies $m_{rad}^2=O(\epsilon)$, which is the
analogue of $m^2=O(\alpha/4\pi)$ in the Coleman-Weinberg mechanism.
For the original GW potential of eqs.~(\ref{vren},\ref{truev}), $P(x)$
is close to having a double zero at $x=v_1/v_0$, which leads to a further
suppression $m_{rad}^2=O(\epsi^{3/2})$. If we modify the model
by adding a contribution
 $\delta T_1=\epsilon v_1^2L^{-4}$ to the IR brane tension, eq.~(\ref{truev})
becomes a perfect square
\beq
V_{new}=(4+2\epsilon)\mu^4(v_1-v_0(\mu/\mu_0)^\epsilon)^2
\label{vcsaki}
\eeq
Now $V_{min}=0$, like in a 
supersymmetric model, while we have exactly 
$\mu_1/\mu_0=(v_1/v_0)^{1/\epsilon}$.
This potential is the one we would have obtained in the
model of ref. \cite{cgk}, by working at leading order in $v^2/(ML)^3$.
The advantage of that model is that one can also easily find the exact
solution of the equations of motions including the backreaction \cite{dewolfe}.
So it is an ideal model to verify the consistency of the effective Lagrangian
approach to the radion potential. By eqs.~(\ref{mass},\ref{vcsaki}) we have
\beq
m_{rad}^2=\frac{v_1^2}{6M^3L^3}(2+\epsilon)\epsilon^2\mu_1^{2}
\eeq	
which agrees with eq. (6.6) of ref. \cite{cgk}.

It is interesting to consider the radion mass from the broader perspective
of a 4d CFT.
As discussed above the slow evolution $d\ln \lambda/d\ln \mu=\epsilon $
leads to a radion that is somewhat lighter than the other modes.
In general, however, the full $\beta$-function will be 
\begin{equation}
\frac{d\ln \lambda}{d\ln \mu}=\epsilon +b_1\lambda+b_2\lambda^2+\dots=\beta(\lambda)
\end{equation}
 and for
large enough $\lambda$ the evolution may become non perturbative  and fast.
In order to generate a hierarchy it is only necessary that $\lambda$ starts
running slowly at the cut-off scale $\mu_0$. Then, if $\lambda$ is irrelevant
($\epsilon>0$) it will keep its slow run until the point where 
$V(\mu)=\mu^4 P(\lambda(\mu))$ is minimized,
and the radion mass will be suppressed. However for the more interesting case
of a relevant ($\epsilon<0$) or marginally relevant ($\epsilon=0$, $b_1<0$
for $\lambda$ positive) coupling the size of $m_{rad}$ depends on whether
$\lambda(\mu)$ is  already running fast at the minimum.
If all the $a_n$ and $b_n$ 
in the expansion of respectively $P$ (see eq. (\ref{colwein}))
and $\beta(\lambda)$ were of order 1 (apart from $b_0=\epsilon$),
then  also $\lambda(\mu)=O(1)$ at the minimum and the radion mass would be unsuppressed.
In a sense this is what happens in technicolor models. Here the deformation
$\lambda$ is a gauge coupling $g^2$ for which $\epsilon=0$ and $b_1<0$, the scale
$\mu$ is precisely the one where $\lambda$ becomes non-perturbative and
there is nothing looking like a light radion in the spectrum.
On the other hand, if, for instance, $a=P(0)$ is small then $P(\lambda)$ can
cross zero for a perturbative $\lambda$ and the radion mass be suppressed.
(This is truly what happens in the Coleman-Weinberg scalar electrodynamics
example \cite{coleman}, where the scalar quartic coupling is playing the role of 
$a$ and the
gauge coupling the role of $\lambda$). This situation must probably be realized
if  $\lambda$ coincides with the top Yukawa coupling $\lambda_t$: the fact that
$\lambda_t$ is perturbative in the low energy theory is evidence
that it is  slowly running in the full CFT just above the weak scale.

\section{\large Two-point functions}
In this Appendix, we compute the two-point function of a minimally
coupled scalar field in the RSI model, corresponding to
a dimension four operator $O$ of the CFT. The same computation also
gives the transverse-traceless part of the two-point function
of the stress-energy tensor. We closely follow
the analogous computation in $AdS$/CFT \cite{witten}.
Put the UV brane at $z=R$ (as a UV regulator) 
and the Tev brane at $z=1/\mu$.
We are interested in the limit where 4d gravity is decoupled
and the UV brane is sent to the boundary $z=0$, therefore 
at the end we will take the limit $R\rightarrow 0$.
Consider the Fourier component $\phi_p$ of a 
minimally coupled massless scalar field. In the
coordinates of eq.~(\ref{e1}), its equation of motion reads 
$(pz)^2\phi^{\prime\prime}-3(pz)\phi^{\prime}-(pz)^2\phi=0$,
whose solution is
\beq
\phi_p(z)=(pz)^2\left [ A(p)K_2(pz)+B(p)I_2(pz)\right ]
\label{b1}
\eeq
In the ordinary $AdS$/CFT computation \cite{witten}, regularity
at the horizon $z=\infty$ selects the solution with $B(p)=0$.
We choose a Neumann condition
$\partial_z\phi(z=1/\mu)=0$ at the Tev brane, 
which fixes $B(p)=A(p)K_1(p/\mu)/I_1(p/\mu)$.  
The value $\phi^0_p=\phi_p(R)$ 
is identified with the boundary source for
the conformal operator $O$. The generating functional $W[\phi^0]$
for the connected Green functions for $O$ can be computed
from supergravity as the bulk action evaluated on the solution
(\ref{b1}) \cite{witten}. As usual, the latter reduces to a boundary
term
\beq
\langle  O(p)O(-p)\rangle ={\partial^2W\over\partial\phi^0_p\partial\phi^0_{-p}}=
\left [{1\over z^4}{\phi_p(z) z\partial_z\phi_p(z)\over\phi_p(R)^2}\right ]^{z=R}_{z=1/\mu}
\label{flux}
\eeq
The contribution at the Tev brane is identically zero due to the boundary
condition. Taking $R\rightarrow 0$, the computation simplifies
\beq
\langle  O(p)O(-p)\rangle =\left [{1\over R^4}z\partial_z\log\phi\right ]^{z=R}=
- {p^4\over 4}\left (\log{pR\over 2}-{K_1(p/\mu)\over
I_1(p/\mu)}\right )
\eeq
The two-point function reduces to the conformal result $-(k^4\log k)/4$
in the UV limit $k\gg \mu$, with exponential corrections,
 and it is analytic for $k\ll \mu$,
where the log is exactly canceled by the Bessel functions. 

The transverse-traceless part of the stress-energy tensor two-point
function as defined in eq.~(\ref{duepunti}) can be computed in terms
of the previous expression,
\beq
{\cal F}(p)=\int d^4p e^{ipx}{c(x)\over x^4}=8\pi^2c{\langle  O(p)O(-p)\rangle \over p^4}
\label{last}\eeq
where the normalization factor has been introduced to match
the UV conformal result ${\cal F}(p)\sim -2\pi^2c\log p$ \cite{witten}.

\def\ijmp#1#2#3{{\it Int. Jour. Mod. Phys. }{\bf #1~}(19#2)~#3}
\def\pl#1#2#3{{\it Phys. Lett. }{\bf B#1~}(19#2)~#3}
\def\zp#1#2#3{{\it Z. Phys. }{\bf C#1~}(19#2)~#3}
\def\prl#1#2#3{{\it Phys. Rev. Lett. }{\bf #1~}(19#2)~#3}
\def\rmp#1#2#3{{\it Rev. Mod. Phys. }{\bf #1~}(19#2)~#3}
\def\prep#1#2#3{{\it Phys. Rep. }{\bf #1~}(19#2)~#3}
\def\pr#1#2#3{{\it Phys. Rev. }{\bf D#1~}(19#2)~#3}
\def\np#1#2#3{{\it Nucl. Phys. }{\bf B#1~}(19#2)~#3}
\def\mpl#1#2#3{{\it Mod. Phys. Lett. }{\bf #1~}(19#2)~#3}
\def\arnps#1#2#3{{\it Annu. Rev. Nucl. Part. Sci. }{\bf #1~}(19#2)~#3}
\def\sjnp#1#2#3{{\it Sov. J. Nucl. Phys. }{\bf #1~}(19#2)~#3}
\def\jetp#1#2#3{{\it JETP Lett. }{\bf #1~}(19#2)~#3}
\def\app#1#2#3{{\it Acta Phys. Polon. }{\bf #1~}(19#2)~#3}
\def\rnc#1#2#3{{\it Riv. Nuovo Cim. }{\bf #1~}(19#2)~#3}
\def\ap#1#2#3{{\it Ann. Phys. }{\bf #1~}(19#2)~#3}
\def\ptp#1#2#3{{\it Prog. Theor. Phys. }{\bf #1~}(19#2)~#3}

\end{document}